\documentclass[12pt]{iopart}
\usepackage{bm}             
\usepackage{graphicx}
\usepackage{float}
\usepackage{amsthm}
\usepackage{amsfonts}
\usepackage{amssymb}
\usepackage{upgreek} 

\begin{document}

\title[Plasmonic Spin-Hall Effect in Ni$_{80}$Fe$_{20}$ microstructures]{Plasmonic Spin-Hall Effect of propagating Surface Plasmon Polaritons in Ni$_{80}$Fe$_{20}$ microstructures}

\author{Maximilian Paleschke$^1$, Cheng-Tien Chiang$^2$, Liane Brandt$^1$, Niklas Liebing$^1$, Georg Woltersdorf$^1$, Wolf Widdra$^1$}

\address{$^1$ Institute of Physics, Martin Luther University Halle-Wittenberg, Von-Danckelmann-Platz 3, 06099 Halle (Saale), Germany}
\address{$^2$ Institute of Atomic and Molecular Sciences, Academia Sinica, Taipei, Taiwan}
\ead{wolf.widdra@physik.uni-halle.de}


\begin{abstract}
	Photoexcitation and shaping of a propagating surface plasmon polariton (SPP) on silver and gold microstructures are well established and lead to the discovery of the plasmonic spin-Hall effect recently. Whereas silver is often the material of choice due to its exceptional low plasma frequency and weak damping, similar observations have not been reported for ferromagnetic metals. In this work, we report on propagating SPPs on Ni$_{80}$Fe$_{20}$ microstructures imaged by photoemission electron microscopy (PEEM) in combination with a tunable femtosecond laser system at MHz repetition rate. Circular dichroic (CD) images in threshold PEEM show clear edge-induced SPPs with sub-micrometer wavelength and propagation length of about 3.5\,$\upmu$m. Analysis of the interference patterns as well as the coupling of the optical spin angular momentum to the observed fringe fields reveal propagation characteristics exclusive to evanescent waves and the presence of the plasmonic spin-Hall effect. Our work provides direct evidence that many materials with a high plasma frequency allow for excitation and observation of propagating SPPs at the dielectric/metal interface via CD PEEM imaging, enabling magnetoplasmonic investigation of common ferromagnets on nanometer length and femtosecond time scales.  
\end{abstract}

\submitto{\NJP}

\maketitle

\section{Introduction}
Due to the remarkable progress in nanofabrication on the one hand and magneto optical methods on the other, the vivid field of magnetoplasmonics has become more and more accessible by a variety of excitation and observation techniques \cite{armelles2013,maccaferri2020}. In general, the term magnetoplasmonics refers to the combination of plasmonic and magnetic properties. Ordinarily, this is achieved within a heterostructure. For instance, typical plasmonic noble metals like silver or gold are used to create plasmons which then interact with magnetic metastructures, films or alloys \cite{alexander2019,bonanni2011,choi2018,luong2019}. 
Another more literal approach to magnetoplasmonics would be to directly excite and utilize plasmons in a magnetic material which couple to the electron and the spin systems. However, the optical properties of most commonly used ferromagnetic materials such as Ni are not ideal and a strong damping of the surface plasmon polariton (SPP) is expected due to its low SPP quality factor \cite{johnson1974}. Whereas the propagation length of an SPP wave in silver can exceed several tens of micrometers \cite{weeber2001,pitarke2007,kahl2018,qin2020}, the amplitude of an SPP is expected to decrease exponentially within a few micrometers in Ni. A second conceptual challenge for the investigation of SPPs at transition metal surfaces arises from the high energy of plasmons. Conventional identification of propagating SPPs and ordinary diffraction phenomena rely on the plasmon dispersion relation which prescribes a distinctly different wavelength of an SPP as compared to that of the incident light \cite{maier2007}. The wavelength of SPPs differs significantly from the free space radiation when their energy is close to the plasma frequency. This plasma frequency is at around 3.7\,eV in silver due to interband transitions contributing to the screening response \cite{dabrowski2020}. But, for most ferromagnetic transition metals it is above 6\,eV, which is higher than the photon energy used in most optical experiments with laboratory light sources \cite{hagelin-weaver2004,bisio2014}. Hence, the corresponding wavelength differences between the SPP wavelength and the wavelength of the incident photon are generally below the spatial resolution of conventional optical microscopy experiments. 

For two electromagnetic waves with very similar energy and momentum, the question might appear as somewhat philosophical: How to experimentally distinguish a plasmon propagating with the velocity of light from an electromagnetic wave propagating in either the metal or the dielectric \cite{pitarke2007}? This question is often circumvented in the community by concentrating on local excitations such as localized surface plasmon polaritons (LSP), which create a strong experimental response via a drastically enhanced photoelectron yield or nonlocal effects via dipolar fields, which are easily detectable even in low resolution conditions. The increase of the photoelectron yield in the vicinity of structured particles or defects has been widely considered as the smoking gun for plasmonic field enhancement and therefore LSPs, and it is extensively investigated for standard plasmonic materials \cite{weeber2001,krenn1999,banfi2003,cinchetti2005} and ferromagnetic metals \cite{bonanni2011,choi2018,sujak2020}. The existence of LSPs, however, implies also the existence of propagating SPPs in these materials. 
In this spirit, another approach coherently launches SPPs with the aid of an array referred to as magnetoplasmonic crystal, resulting in a plasmonic standing wave pattern and a resonant enhancement of the magneto-optical activity \cite{maccaferri2015,rollinger2016}. Following this approach in photoemission electron microscopy (PEEM) experiments \cite{rollinger2016}, SPPs at magnetic microstructures were imaged directly with nanometer spatial resolution using variable incident light polarization and photon energy. 

\begin{figure}[H]
	\includegraphics*[width=\linewidth]{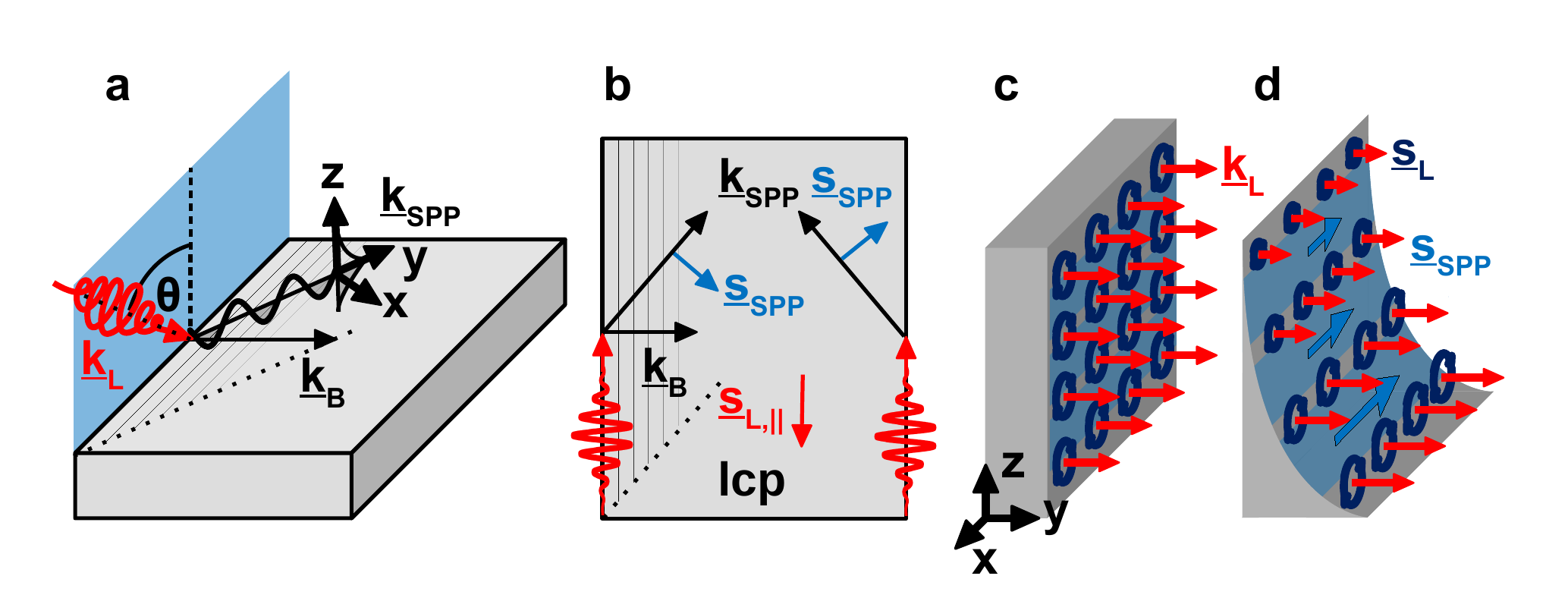}
	\caption{\textbf{(a)} Schematics of SPP excitation by a circular polarized laser pulse at the edge of a Ni$_{80}$Fe$_{20}$ microstructure. The laser pulse excites the microstructure under an angle of incidence $\Theta$ with respect to the surface normal. At the edge, a surface plasmon polariton is launched, which propagates in a direction (y direction) defined by the phase matching based on the surface refractive index. Interference of the incoming laser plane wave with the propagating plasmon field results in a modulation of the PEEM intensity parallel to the edge as detected in the experiment (gray lines). The plasmon propagation direction leads to a shadowing area of the interference pattern starting at the corner of the microstructure (black dotted line parallel to the y direction). \textbf{(b)} Schematics of the plasmonic spin-Hall effect, where the SAM of the incoming circularly polarized light couples to the parallel component of the SAM of the SPP, leading to an enhanced excitation on the left edge for lcp photoexcitation. \textbf{(c)} Schematic cut through a freely propagating light wave front with circular polarization. \textbf{(d)} Similar schematic for an evanescent wave, revealing the angular momentum $\vec{s}_{SPP}$ when integrating over the blue-shaded area due to the field gradient in z-direction.}
	\label{fig:cartoon}
\end{figure}

In the present study, we demonstrate a straightforward approach to investigate propagating SPPs launched at magnetic microstructures, utilizing circular dichroism (CD) imaging in PEEM. 
Simultaneously, we use the evanescent character of propagating SPPs to differentiate the bound surface state from free space radiation often present in PEEM measurements \cite{chelaru2006,marchenko2011,tavassoly2009,word2016}. It has been recently discovered that the evanescent electric field creates an additional spin angular momentum (SAM) perpendicular to the propagation direction of the SPP \cite{bliokh2012,bliokh2014,bliokh2015}. Experimentally, this additional SAM allows for selective excitation of SPPs at different edges of a microstructure as shown in figure \ref{fig:cartoon} \cite{lin2013,dai2018}, and can be related to the plasmonic spin-Hall effect \cite{dai2019}. Since this effect relies on the presence of an additional SAM to which the SAM of the incident light can be coupled, it is not present in a plane wave propagating in the bulk \cite{bliokh2014}. Such coupling has already been discovered experimentally on patterned silver samples. As a significant extension of the existing plasmonics of noble metals, we will show in the following that the plasmonic spin-Hall effect is also present in Ni$_{80}$Fe$_{20}$ (permalloy) microstructures and can serve as a tool to directly characterize propagating SPPs at magnetic surfaces. 

We demonstrate that the plasmonic spin-Hall effect in Ni$_{80}$Fe$_{20}$ microstructures leads to CD in threshold laser PEEM imaging. We facilitate this observation to identify propagating SPPs on Ni$_{80}$Fe$_{20}$ for the first time and extract their wavelength and propagation length, providing valuable insights into the dielectric properties at the surface.  

\section{Experimental details}

Microstructures of metallic Ni$_{80}$Fe$_{20}$ were defined on undoped GaAs(001) substrates using electron beam lithography, thermal evaporation and lift-off. The morphology of the 30\,nm thick polycrystalline Ni$_{80}$Fe$_{20}$ microstructures was verified \textit{ex situ} via atomic force microscopy (AFM). The magnetic domain pattern was examined \textit{in situ} via magneto-optic Kerr effect (MOKE) microscopy and \textit{ex situ} via magnetic-force microscopy (MFM).

A 20\,W fiber laser (Impulse, Clark-MXR, Dexter, USA) with a repetition rate of 1.23\,MHz was used to pump a noncollinear optical parametric amplifier (NOPA) and to generate femtosecond laser pulses with a widely tunable photon energy range \cite{Duncker12,Gillmeister18,gillmeister2020}. Here, frequency-doubling of the NOPA output is used for photon energies of 3.32\,eV and 4.51\,eV with pulse lengths of approximately 40\,fs. The latter photon energy results in a dominant one-photon photoemission (1PPE) process leading to a high photoelectron yield at the microstructures as compared to the substrate. Photoexcitation at 3.32\,eV leads to an inverted PEEM contrast between the Ni$_{80}$Fe$_{20}$ structure and the GaAs substrate due to the dominant excitation of electrons from the valence bands of GaAs via a two-photon photoemission (2PPE) process. These 2PPE and 1PPE processes were verified by variation of the pump power, which revealed a non-linear and a linear increase of the photoelectron yield, respectively. As a result of residual gas adsorption after growth and during sample transfer, the measured 1PPE threshold lies below the known work function of clean Ni$_{80}$Fe$_{20}$ of 5.0\,eV \cite{salou2008, weiss2002}. The laser-induced photoconductivity of the otherwise undoped GaAs substrate helped to suppress charging effects in PEEM \cite{huber2001,ishioka2015}.

The measurements were performed under ultra-high vacuum conditions at a base pressure of about $5\times10^{-10}$\,mbar with a PEEM (FOCUS GmbH, Hünstetten, Germany). Prior to every measurement the spatial resolution was optimized to around 20\,nm via calibration measurements with a mercury discharge lamp. The laser beam polarization was controlled with appropriate achromatic quarter-wave plates, which were automatically rotated via a piezoelectric rotation mount for fast polarization switching. Space charge effects were avoided by reducing the pulse power to 1\,nJ. In addition, the measurement for each polarization direction was accumulated for at least 30\,min to obtain a sufficient signal-to-noise ratio. To enhance the quality of the dichroic signals, the light helicity was automatically alternated every 10\,s in order to suppress long-term drifts of the laser intensity. The laser spot size on the sample can be varied down to a diameter of 10\,$\upmu$m via a movable parabolic mirror, and was suitably chosen to illuminate the field of view as well as to avoid space charge effects. 

Figure \ref{fig:cartoon}(a) shows a sketch of the excitation wave vectors in the experimental configuration as well as the influence of the evanescent character of an SPP compared to free space radiation. Optical excitation of an SPP requires to overcome the parallel momentum mismatch at a given energy. The additional momentum is typically provided by a notch or an edge on the sample \cite{kubo2007}. At such a discontinuity, the SPP is launched by an ultrashort laser pulse and travels along the metal-dielectric interface. Due to the light angle of incidence with respect to the surface normal $\Theta = 65^{\circ}$ in our experiment, the pump laser pulse interferes with the SPP launched, giving rise to a plasmonically enhanced two-photon photoemission yield \cite{kubo2007}.

In order to separate the aforementioned plasmonic spin-Hall effect from other contrast mechanisms in PEEM, we apply helicity-dependent measurements and evaluate the subsequent dichroic PEEM intensities. The dichroic signals are defined quantitatively as follows:

\begin{equation}
	A = \frac{I_{lcp} - I_{rcp}}{I_{lcp} + I_{rcp}}\quad .
	\label{eq:dichroism}
\end{equation}

Here, $I_{lcp}$ and $I_{rcp}$ stand for the PEEM intensities measured for excitation with left-circularly polarized (lcp) and right-circularly polarized light (rcp), respectively. A helicity dependent signal will result in a non-vanishing asymmetry value $A$ contributing to the dichroic PEEM image. As depicted in figure \ref{fig:cartoon}(b), the plasmonic spin-Hall effect will cause lcp light to predominantly excite photoelectrons on the left edge of the structure in PEEM, whilst rcp enhances photoemission from the right edge. Hence, this plasmonic spin-Hall helicity dependence results in a dichroic interference pattern in the PEEM image, where the phase of the oscillation depends on the direction of the edge. As will be experimentally shown, calculating the asymmetry according to equation \ref{eq:dichroism} results in a phase shift of $\pi$ when comparing the interference patterns at opposite edges.

\section{Results and Discussion}

\subsection{SPP propagation length}
\begin{figure}[H]
	\includegraphics*[width=\linewidth]{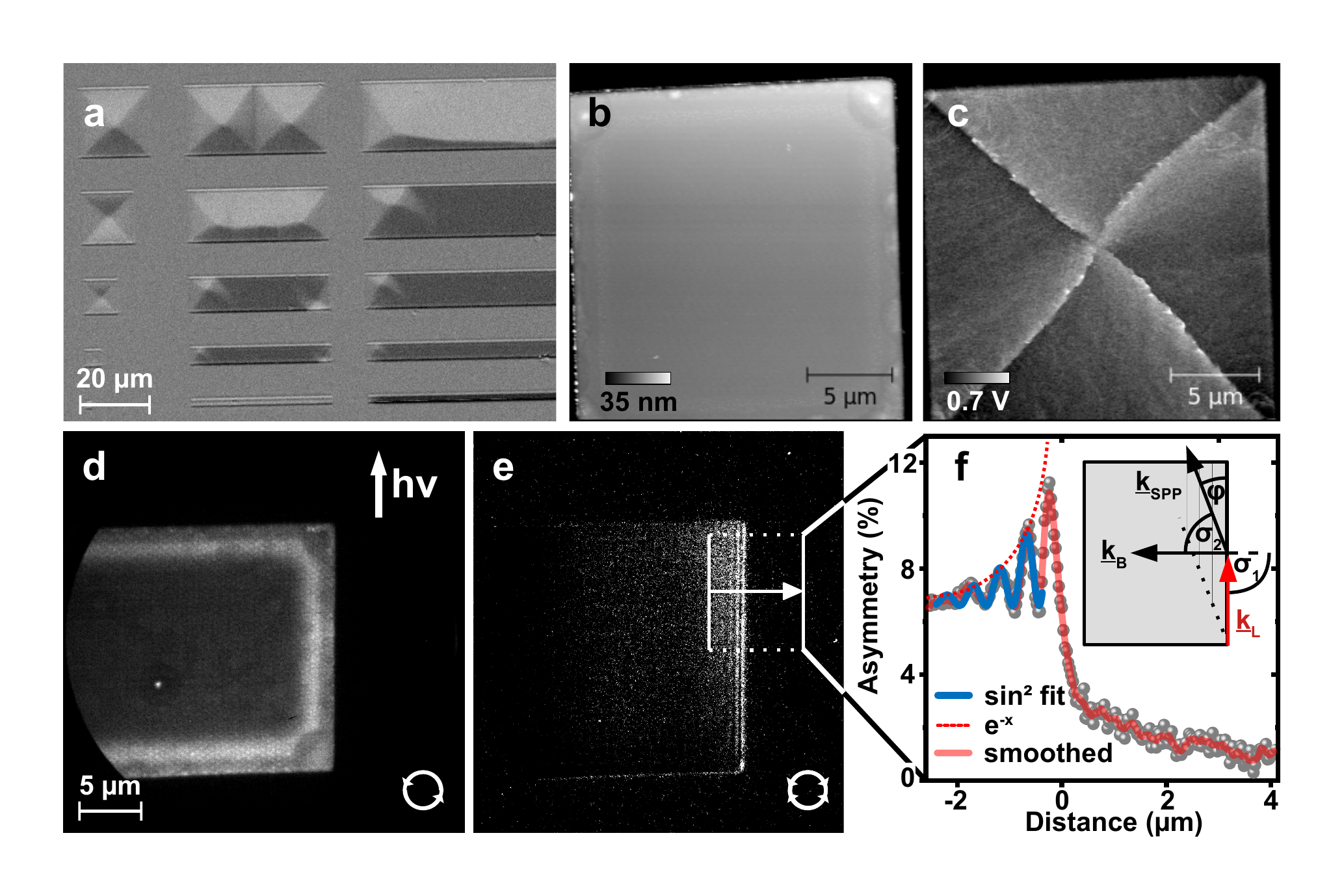}
	\caption{\textbf{(a)} MOKE image of Ni$_{80}$Fe$_{20}$ microstructures grown on GaAs(001). Magnetic domains of typical Landau patterns are visible. An AFM \textbf{(b)} and MFM \textbf{(c)} image of the square in the top-left corner in (a) displaying the height and domain walls. \textbf{(d)} Laser PEEM image of the rectangular Ni$_{80}$Fe$_{20}$ island in the middle column of (a) for lcp light with 4.51\,eV, with the direction of light incidence marked in the upper right. \textbf{(e)} CD PEEM image of the same structure in (d) with the interference pattern at the right edge of the Ni$_{80}$Fe$_{20}$ microstructure. \textbf{(f)} Line scan of the region marked in (e) with an inset showing the wave vectors $k_L$ and $k_{SPP}$.}
	\label{fig:Py_overview}
\end{figure}

Figure \ref{fig:Py_overview} shows the Ni$_{80}$Fe$_{20}$ microstructures grown on GaAs with the MOKE image in panel (a) showing an array of Ni$_{80}$Fe$_{20}$ squares and rectangles. The clear magnetic contrast reveals the expected Landau domain pattern \cite{landau1935}. The surface morphology and stability of the domain pattern was verified via AFM and MFM as shown in panel (b) and (c), respectively. The AFM image demonstrates the overall flatness of the Ni$_{80}$Fe$_{20}$ structures with an average thickness of 30\,nm and a small decrease of approximately 1\,nm in height at the edges. Figure \ref{fig:Py_overview}(d) shows a laser excited PEEM image of one rectangular Ni$_{80}$Fe$_{20}$ island for photoexcitation with lcp light at a photon energy of 4.51\,eV. Here, the dominant mechanisms contributing to the PEEM contrast are the sample topography and the difference in the surface work function of substrate and metallic structure \cite{schneider2002}. The CD image derived according to equation \ref{eq:dichroism} is shown in panel (e), revealing an interference pattern in the vicinity of the edge. It can be seen more clearly in the line scan in figure \ref{fig:Py_overview}(f). In both panels (d) and (e), no magnetic contrast is visible. 

As has been previously demonstrated for the plasmonic materials silver and gold, the wavelength and exponential decay of the emerging interference pattern $\lambda_M$ can be used to estimate the SPPs wavelength $\lambda_{SPP}$ and its propagation length $L_{SPP}$ \cite{maier2007, creath92, buckanie2013}. The SPP wavelength can be extracted from the equation 

\begin{equation}
	\lambda_M = \frac{\lambda_{L,\vert\vert} \lambda_{SPP}}{\sqrt{\lambda^2_{L,\vert\vert} + \lambda_{SPP}^2 - 2\lambda_{L,\vert\vert}\lambda_{SPP}\cdot\cos{\phi}}}\quad ,
	\label{eq:moire_pattern}
\end{equation}

where $\lambda_{L,\vert\vert} = \frac{\lambda_L}{\sin\Theta}$ refers to the wavelength component of the incident light parallel to the surface. The angle $\phi$ is spanned by the vectors $\vec{k}_L$ and $\vec{k}_{SPP}$ and can be derived experimentally from the angle between the edge of the microstructure and the onset of the interference pattern observed in the PEEM image as marked by the dotted line in the inset of figure \ref{fig:Py_overview}(f) or in figure \ref{fig:cartoon}(a) and (b). Alternatively, $\phi$ can be calculated by Snell's law according to

\begin{equation}
	\phi = \sigma_1 - \sigma_2 = \sigma_1 - \arcsin\left(\frac{n_1}{n_2}\sin\sigma_1\right)\quad ,
	\label{eq:snell} 
\end{equation}
where $\sigma_1$ refers to the angle between the projection of the incident light wave vector and the normal vector of the edge where the SPP is launched. $\sigma_2$ describes the angle between the edge normal and the direction of $\vec{k}_{SPP}$, and $n_1$ and $n_2$ are the refractive indices of the dielectric and the metal surface, respectively. Therefore, $\lambda_{SPP}$ can be extracted from the observed interference pattern according to the experimental parameters. Furthermore, $\lambda_{SPP}$ is fully determined by the complex dielectric function of the materials at the incident photon energy. When we take $n_1=1$ for vacuum, the experimentally observed angle of $\phi = 12\,^{\circ}\pm5\,^{\circ}$ corresponds to a surface refractive index of $n_2 = 1.02 \pm 0.02$ for Ni$_{80}$Fe$_{20}$ at 3.32\,eV. This value agrees well with the tabulated bulk value for Ni$_{80}$Fe$_{20}$ of $n_2 = 1.017$ \cite{tikuisis2017}, where 
\begin{equation}
	n_{\mathrm{surface}} = \mathrm{Re}\left(\sqrt{\frac{\epsilon_{\mathrm{bulk}}}{1+\epsilon_{\mathrm{bulk}}}}\right)\quad .
	\label{eq:surface_n}
\end{equation}

Fitting an exponentially damped sine squared function to the observed interference pattern provides an estimation for the SPP propagation length $L_{SPP}$, where

\begin{equation}
	L_{SPP} = \frac{L_M}{\sin\phi} = k_0 \left(\frac{1+\mathrm{Re}(\epsilon)}{\mathrm{Re}(\epsilon)}\right)^{3/2}\frac{\left(\mathrm{Re}(\epsilon)\right)^2}{\mathrm{Im}(\epsilon)}\quad .
	\label{eq:prop_length}
\end{equation}

Here, the complex dielectric function of the metal is denoted by $\epsilon$, $k_0$ is the wave vector of the incident light in vacuum and $L_M$ is the decay length of the interference pattern \cite{maier2007}. An example for the intensity decay of the interference pattern is shown in figure \ref{fig:Py_overview}(f), from which a propagation length of $L_{SPP, Py} = \left(3.45\pm1.01\right)$\,$\upmu$m for Ni$_{80}$Fe$_{20}$ at 3.32\,eV can be derived. The large uncertainty results mainly from the relative error of $\phi$. The value of $L_{SPP, Py}$ from the experiments agrees well with the theoretical value based on the bulk dielectric function of Ni$_{80}$Fe$_{20}$ \cite{tikuisis2017}, which is $L_{SPP, Py} \approx 3.53$\,$\upmu$m. By using a laser pulse duration of 40\,fs, the SPP propagation length is experimentally accessible without the need of a pump-probe delay up to about 13\,$\upmu$m, which is much longer than $L_{SPP, Py}$ observed here.

To confirm our analysis, we additionally investigated propagating SPPs for silver microstructures on Si(001) by PEEM (data not shown). For Ag, we find an experimental propagation length of $L_{SPP, Ag} = \left(1.20\pm0.33\right)$\,$\upmu$m at a photon energy of 3.32\,eV, which compares well with the theoretical value of $L_{SPP, Ag} \approx 1.21$\,$\upmu$m based on known optical constants of Ag thin films \cite{ciesielski2017}. The good agreement in both cases corroborates our approach.

\begin{figure}[H]
	\includegraphics*[width=\linewidth]{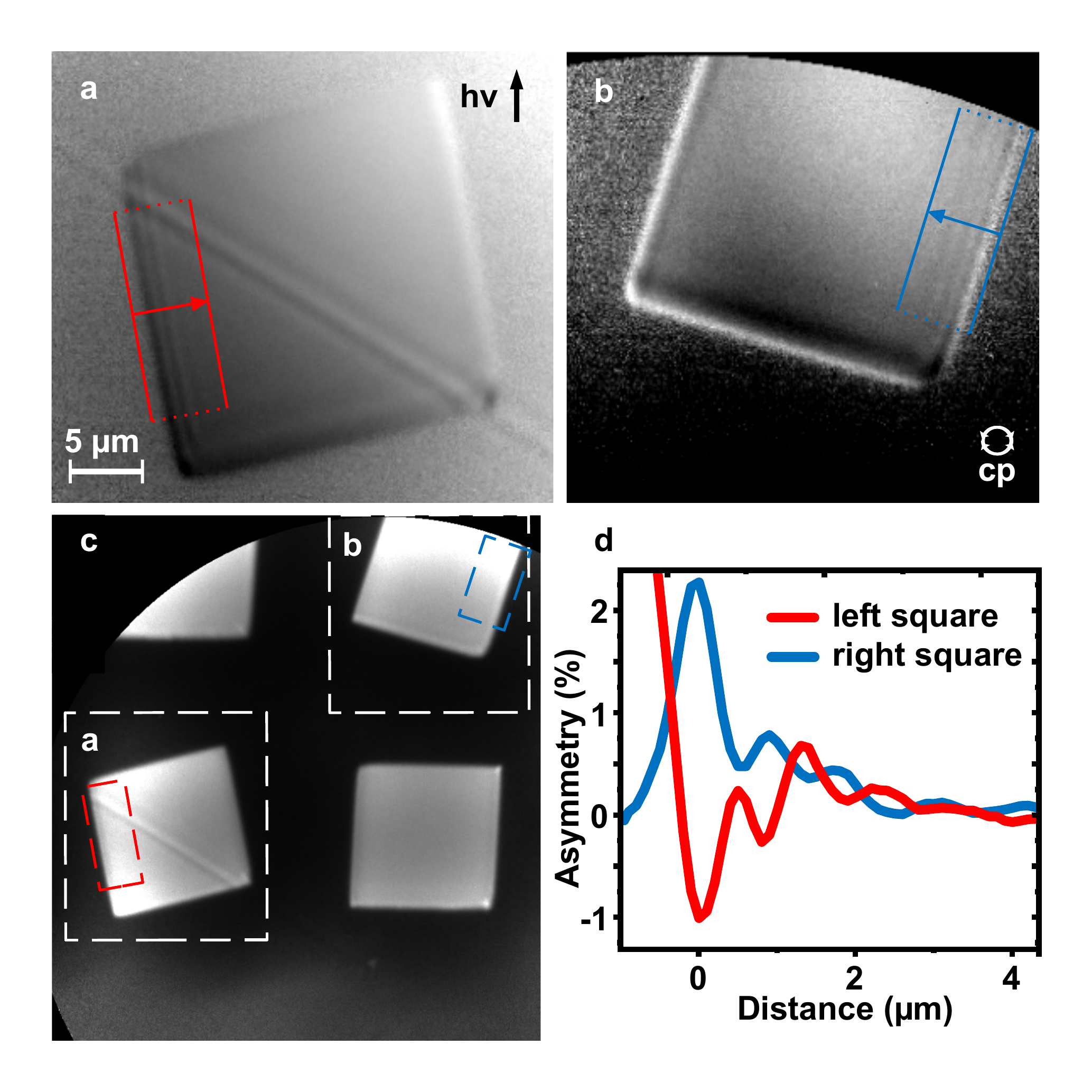}
	\caption{CD PEEM images of rotated Ni$_{80}$Fe$_{20}$ microstructures \textbf{(a)} and \textbf{(b)} measured at the incident laser photon energy of 4.51\,eV showing edge-induced interference patterns, with the direction of light incidence shown in the upper right of panel (a). These structures are marked in the overview PEEM image \textbf{(c)} measured at a photon energy of 5.2\,eV with a Hg discharge lamp. The aligned line scans in \textbf{(d)} of the region marked in (a) and (b) reveal their identical wavelength and a phase shift of $\pi$.}
	\label{fig:1PPE_plasmon_2}
\end{figure}

Comparison of the dichroic interference patterns of Ni$_{80}$Fe$_{20}$ structures that are tilted in-plane by $\pm\,15^{\circ}$ with respect to the direction of the incident light in figure \ref{fig:1PPE_plasmon_2}(a,b) reveals a longer $\lambda_M$ than in figure \ref{fig:Py_overview} for the none-rotated island. This observation can be attributed to a smaller $\phi$ according to equation \ref{eq:moire_pattern}. Additionally, the interference is visible only at the left edge for the Ni$_{80}$Fe$_{20}$ island rotated by $-15^{\circ}$ in figure \ref{fig:1PPE_plasmon_2}(a) and at the right edge for $+15^{\circ}$ in figure \ref{fig:1PPE_plasmon_2}(b). This phenomenon can be explained by Snell's law, since $\sigma_1 > 90^{\circ}$ at the edges where no interference can be observed. Line scans of the interference patterns on both Ni$_{80}$Fe$_{20}$ islands show a clear phase shift of $\pi$ as depicted in figure \ref{fig:1PPE_plasmon_2}(d) with an otherwise similar amplitude and damping. This phase shift will be explained in the next section.

\newpage
\subsection{Plasmonic spin-Hall effect in CD PEEM}

\begin{figure}[H]
	\includegraphics*[width=\linewidth]{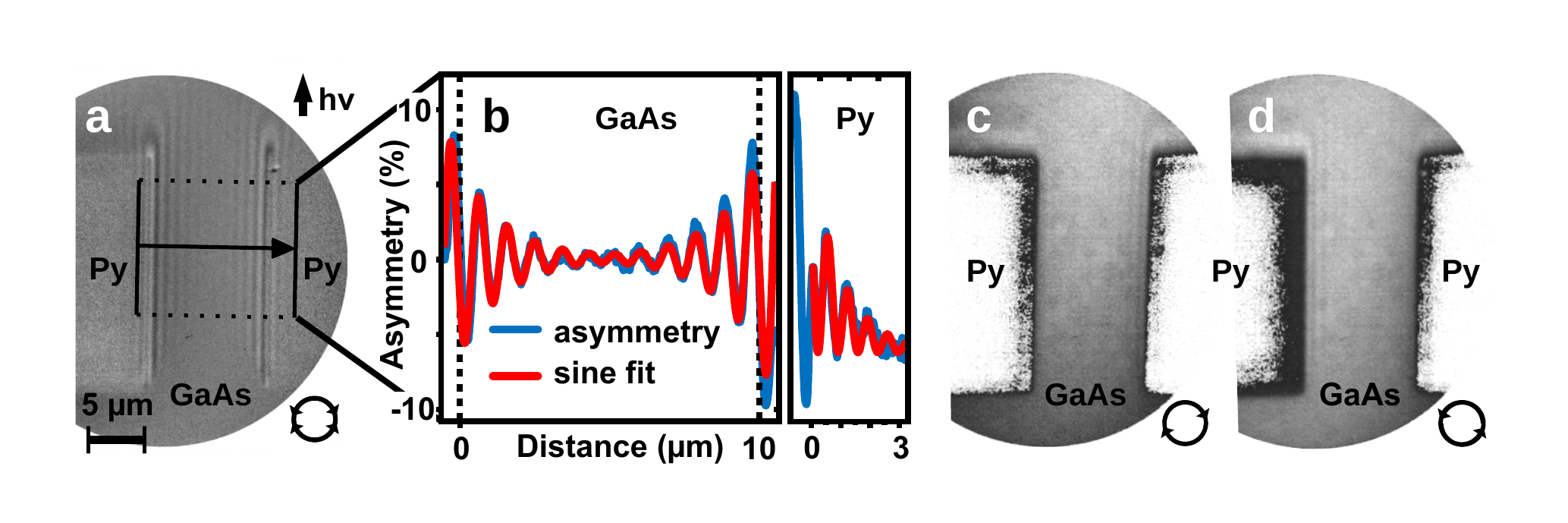}
	\caption{\textbf{(a)} CD PEEM image of two Ni$_{80}$Fe$_{20}$ microstructures and the GaAs substrate region in between measured with a photon energy of 3.32\,eV. The direction of light incidence is marked in the upper right. \textbf{(b)} CD line scan of the marked gap area in (a) and fit by the superposition of two damped sinusoidal waves with constant wavelength (red). The positions of the edges are marked in (b) by  vertical dashed lines. \textbf{(c)} and \textbf{(d)} show PEEM images for rcp and lcp incident light after background subtraction, respectively.}
	\label{fig:cp_overview2}
\end{figure}

According to theory, the plasmonic spin-Hall effect should be independent of the photon energy of excitation \cite{bliokh2012,bliokh2014,bliokh2015}. Indeed, we find SAM-dependent excitation of SPPs also at 3.32\,eV as depicted in figure \ref{fig:cp_overview2}. By changing the excitation energy from 4.51\,eV to 3.32\,eV, the PEEM contrast mechanism changes from the 1PPE to the 2PPE process. The 2PPE process allows us to excite SPPs and observe the resulting interference patterns on Ni$_{80}$Fe$_{20}$ as well as on the photodoped GaAs. Figure \ref{fig:cp_overview2}(a) shows the CD PEEM image for 3.32\,eV excitation on two rectangular Ni$_{80}$Fe$_{20}$ microstructures with the bare GaAs substrate surface in between, where a clear SPP interference pattern is observed. The interference pattern on the photodoped GaAs has a higher amplitude and a longer wavelength than that on Ni$_{80}$Fe$_{20}$ and can be attributed to its different surface refractive index. Figure \ref{fig:cp_overview2}(b) shows the line scan from one Ni$_{80}$Fe$_{20}$ microstructure to the other across the GaAs gap. The asymmetry signal reveals two exponentially damped sinusoidal waves of constant wavelength starting at each edge of the Ni$_{80}$Fe$_{20}$ microstructures. The superposition of both sinusoidal waves in the GaAs region results in an antisymmetric pattern with respect to the gap center as fitted by the solid red line. Note that the antisymmetry of the wave pattern corresponds directly to the $\pi$ phase shift observed in figure \ref{fig:1PPE_plasmon_2}(d), which manifests the plasmonic spin-Hall effect.   

The strong signal in the gap region allows to clarify the origin of the dichroism by analyzing the raw PEEM images for each light polarization. After subtraction of a smooth background, the individual PEEM image for rcp and lcp is depicted in figure \ref{fig:cp_overview2}(c) and (d), respectively. Figure \ref{fig:cp_overview2}(c) shows the interference pattern and SPP excitation mainly at the right edge, whereas the SPP in figure \ref{fig:cp_overview2}(d) is observed at the left edge. This edge-selective SPP excitation is characteristic for the plasmonic spin-Hall effect. 

\begin{figure}[H]
	\includegraphics*[width=\linewidth]{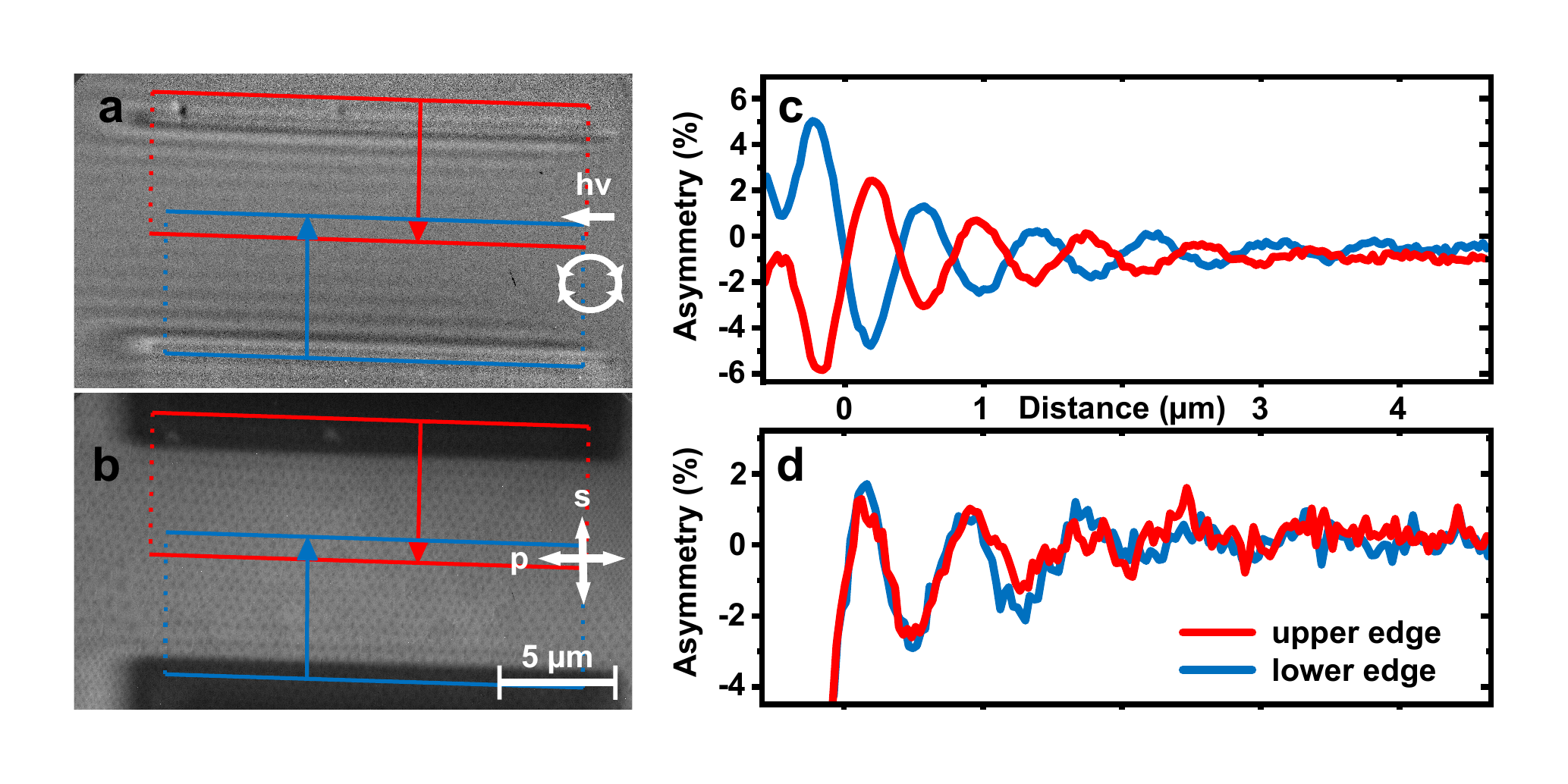}
	\caption{Circular versus linear dichroism in laser PEEM images of the region between two Ni$_{80}$Fe$_{20}$ microstructures measured at the incident photon energy of 3.32\,eV in \textbf{(a)} and \textbf{(b)}, respectively. The direction of light incidence is from right to left. Dichroic line scans are depicted in \textbf{(c)} and \textbf{(d)} starting from both edges for circular and linear polarization, respectively. For the CD in (c), a phase shift of  $\pi$ is observed when comparing the dichroic signals of the upper and lower edges. This phase shift is absent for the linear dichroism in (d).}
	\label{fig:cp_vs_lp}
\end{figure}

To additionally support that the observed phase shift of $\pi$ in CD PEEM is a result of the plasmonic spin-Hall effect, we compare images obtained with linearly and circularly polarized light. Figure \ref{fig:cp_vs_lp}(a) shows a CD PEEM image of the GaAs gap with the line scans in figure \ref{fig:cp_vs_lp}(c) of the marked areas, with the latter revealing the aforementioned phase shift of $\pi$. Figure \ref{fig:cp_vs_lp}(b) shows the linear dichroic image of the same region and the corresponding line scans in figure \ref{fig:cp_vs_lp}(d). In the latter images, the same wavelength of the interference patterns as in figure \ref{fig:cp_vs_lp}(a) is observed but without any significant phase shift. This observation is consistent with the theoretical considerations explained above as well as previous experiments on silver \cite{dai2019}. The interference in the linear dichroic signal originates dominantly from s-polarization, because p-polarized light has no wave vector component perpendicular to the edge of the structure. Hence, when calculating the asymmetry according to equation \ref{eq:dichroism} interchanging $I_{lcp}$ and $I_{rcp}$ with $I_{s}$ and $I_{p}$, respectively, a dichroic signal remains without a phase shift when comparing both edges.

\newpage
\section{Conclusion}

In this work, we analyzed the circular and linear dichroism in threshold PEEM images of Ni$_{80}$Fe$_{20}$ microstructures on GaAs. Interference patterns surrounding their edges reveal the plasmonic spin-Hall effect. Detailed analysis of the interference patterns provides an estimation for the surface refractive index and the SPP propagation length. Moreover, a characteristic phase shift of $\pi$ is observed in the CD PEEM images for the SPP launched at opposite edges, which is a direct consequence of the plasmonic spin-Hall effect. We conclude that the observed interference patterns directly result from excitation of SPPs on Ni$_{80}$Fe$_{20}$ and can be used to quantify the surface refractive index. Our results can be further extended to many magnetic materials for the study of SPP phenomena. Since we do not measure any magnetic signal in CD PEEM for Ni$_{80}$Fe$_{20}$, we can not conclude on the sensitivity of the plasmonic spin-Hall effect to the magnetization direction. In the non-linear regime however, results on magnetoplasmonic multilayers demonstrate strong magnetization-sensitive effects \cite{temnov2010,zheng2015}. Incorporating these approaches with pure ferromagnetic metals would allow for exploration of interesting interplay between magnetic and plasmonic properties on nanometer and femtosecond spatial-temporal scales. 

\ack
This work was supported by the German Research Foundation (DFG) within the SFB/TRR 227 Ultrafast Spin Dynamics (projects A06 and B01). Additionally, we thank R. Kulla for technical support during setup of the experiment.

\section*{References}
\bibliographystyle{iopart-num}
\bibliography{bibtex_bibliography_short}

\providecommand{\newblock}{}
\begin{thebibliography}{10}
\expandafter\ifx\csname url\endcsname\relax
  \def\url#1{{\tt #1}}\fi
\expandafter\ifx\csname urlprefix\endcsname\relax\def\urlprefix{URL }\fi
\providecommand{\eprint}[2][]{\url{#2}}

\bibitem{armelles2013}
Armelles G, Cebollada A, {Garc{\'i}a-Mart{\'i}n} A and Gonz{\'a}lez M~U 2013
  {\em Advanced Optical Materials\/} {\bf 1} 10--35

\bibitem{maccaferri2020}
Maccaferri N, Zubritskaya I, Razdolski I, Chioar I~A, Belotelov V, Kapaklis V,
  Oppeneer P~M and Dmitriev A 2020 {\em Journal of Applied Physics\/} {\bf 127}
  080903

\bibitem{alexander2019}
Alexander D~T~L, Forrer D, Rossi E, Lidorikis E, Agnoli S, Bernasconi G~D,
  Butet J, Martin O~J~F and Amendola V 2019 {\em Nano Letters\/} {\bf 19}
  5754--5761

\bibitem{bonanni2011}
Bonanni V, Bonetti S, Pakizeh T, Pirzadeh Z, Chen J, Nogu{\'e}s J, Vavassori P,
  Hillenbrand R, {\AA}kerman J and Dmitriev A 2011 {\em Nano Letters\/} {\bf
  11} 5333--5338

\bibitem{choi2018}
Choi B~C, Xu H, Hajisalem G and Gordon R 2018 {\em Applied Physics Letters\/}
  {\bf 112} 022403

\bibitem{luong2019}
Luong H~M, Pham M~T, Ai B, Nguyen T~D and Zhao Y 2019 {\em Physical Review B\/}
  {\bf 99} 224413

\bibitem{johnson1974}
Johnson P and Christy R 1974 {\em Physical Review B\/} {\bf 9} 5056--5070

\bibitem{weeber2001}
Weeber J~C, Krenn J~R, Dereux A, Lamprecht B, Lacroute Y and Goudonnet J~P 2001
  {\em Physical Review B\/} {\bf 64} 045411

\bibitem{pitarke2007}
Pitarke J, Silkin V, Chulkov E and Echenique P 2007 {\em Reports on Progress in
  Physics\/} {\bf 70} 1

\bibitem{kahl2018}
Kahl P, Podbiel D, Schneider C, Makris A, Sindermann S, Witt C, Kilbane D,
  {Horn-von Hoegen} M, Aeschlimann M and {zu Heringdorf} F~J~M 2018 {\em
  Plasmonics\/} {\bf 13} 239--246

\bibitem{qin2020}
Qin Y, Ji B, Song X and Lin J 2020 {\em Photonics Research\/} {\bf 8} 1042

\bibitem{maier2007}
Maier S~A 2007 {\em Plasmonics: Fundamentals and Applications\/} ({New York}:
  {Springer})

\bibitem{dabrowski2020}
D{\c a}browski M, Dai Y and Petek H 2020 {\em Chemical Reviews\/} {\bf 120}
  6247--6287

\bibitem{hagelin-weaver2004}
Hagelin-Weaver H~A, Weaver J~F, Hoflund G~B and Salaita G~N 2004 {\em Journal
  of Electron Spectroscopy and Related Phenomena\/} {\bf 134} 139--171

\bibitem{bisio2014}
Bisio F, Proietti~Zaccaria R, Moroni R, Maidecchi G, Alabastri A, Gonella G,
  Giglia A, Andolfi L, Nannarone S, Mattera L and Canepa M 2014 {\em ACS
  Nano\/} {\bf 8} 9239--9247

\bibitem{krenn1999}
Krenn J~R, Dereux A, Weeber J~C, Bourillot E, Lacroute Y, Goudonnet J~P,
  Schider G, Gotschy W, Leitner A, Aussenegg F~R and Girard C 1999 {\em
  Physical Review Letters\/} {\bf 82} 2590--2593

\bibitem{banfi2003}
Banfi G, Ferrini G, Peloi M and Parmigiani F 2003 {\em Physical Review B\/}
  {\bf 67} 035428

\bibitem{cinchetti2005}
Cinchetti M and Sch{\"o}nhense G 2005 {\em Journal of Physics: Condensed
  Matter\/} {\bf 17} S1319--S1328

\bibitem{sujak2020}
Sujak M and Djuhana D 2020 {\em Key Engineering Materials\/} {\bf 855} 243--247

\bibitem{maccaferri2015}
Maccaferri N, Inchausti X, {Garc{\'i}a-Mart{\'i}n} A, Cuevas J~C, Tripathy D,
  Adeyeye A~O and Vavassori P 2015 {\em ACS Photonics\/} {\bf 2} 1769--1779

\bibitem{rollinger2016}
Rollinger M, Thielen P, Melander E, {\"O}stman E, Kapaklis V, Obry B, Cinchetti
  M, {Garc{\'i}a-Mart{\'i}n} A, Aeschlimann M and Papaioannou E~T 2016 {\em
  Nano Letters\/} {\bf 16} 2432--2438

\bibitem{chelaru2006}
Chelaru L~I, {Horn-von Hoegen} M, Thien D and {zu Heringdorf} F~J~M 2006 {\em
  Physical Review B\/} {\bf 73} 115416

\bibitem{marchenko2011}
Marchenko P, Orlov S, Huber C, Banzer P, Quabis S, Peschel U and Leuchs G 2011
  {\em Optics Express\/} {\bf 19} 7244 (\textit{Preprint} \eprint{1102.2140})

\bibitem{tavassoly2009}
Tavassoly M~T, Amiri M, Darudi A, Aalipour R, Saber A and Moradi A~R 2009 {\em
  Journal of the Optical Society of America A\/} {\bf 26} 540

\bibitem{word2016}
Word R~C, Fitzgerald J and K{\"o}nenkamp R 2016 {\em Ultramicroscopy\/} {\bf
  160} 84--89

\bibitem{bliokh2012}
Bliokh K~Y and Nori F 2012 {\em Physical Review A\/} {\bf 85} 061801

\bibitem{bliokh2014}
Bliokh K~Y, Bekshaev A~Y and Nori F 2014 {\em Nature Communications\/} {\bf 5}
  3300

\bibitem{bliokh2015}
Bliokh K~Y, {Rodr{\'i}guez-Fortu{\~n}o} F~J, Nori F and Zayats A~V 2015 {\em
  Nature Photonics\/} {\bf 9} 796--808

\bibitem{lin2013}
Lin J, Mueller J~P~B, Wang Q, Yuan G, Antoniou N, Yuan X~C and Capasso F 2013
  {\em Science\/} {\bf 340} 331--334

\bibitem{dai2018}
Dai Y, Dabrowski M, Apkarian V~A and Petek H 2018 {\em ACS Nano\/} {\bf 12}
  6588--6596

\bibitem{dai2019}
Dai Y and Petek H 2019 {\em ACS Photonics\/} {\bf 6} 2005--2013

\bibitem{Duncker12}
Duncker K, Kiel M and Widdra W 2012 {\em Surface Science\/} {\bf 606} L87--L90

\bibitem{Gillmeister18}
Gillmeister K, Kiel M and Widdra W 2018 {\em Physical Review B\/} {\bf 97}
  085424

\bibitem{gillmeister2020}
Gillmeister K, Gole{\v z} D, Chiang C~T, Bittner N, Pavlyukh Y, Berakdar J,
  Werner P and Widdra W 2020 {\em Nature Communications\/} {\bf 11} 4095

\bibitem{salou2008}
Salou M, Lescop B, Rioual S, Lebon A, Youssef J~B and Rouvellou B 2008 {\em
  Surface Science\/} {\bf 602} 2901--2906

\bibitem{weiss2002}
Weiss W and Ranke W 2002 {\em Progress in Surface Science\/}  151

\bibitem{huber2001}
Huber R, Tauser F, Brodschelm A, Bichler M, Abstreiter G and Leitenstorfer A
  2001 {\em Nature\/} {\bf 414} 286--289

\bibitem{ishioka2015}
Ishioka K, Brixius K, H{\"o}fer U, Rustagi A, Thatcher E~M, Stanton C~J and
  Petek H 2015 {\em Physical Review B\/} {\bf 92} 205203

\bibitem{kubo2007}
Kubo A, Pontius N and Petek H 2007 {\em Nano Letters\/} {\bf 7} 470--475

\bibitem{landau1935}
Landau L and Lifshits E 1935 {\em Phys. Zeitsch. der Sow.\/} {\bf 8} 153–169

\bibitem{schneider2002}
Schneider C~M and {Schönhense} G 2002 {\em Reports on Progress in Physics\/}
  {\bf 65} 1785--1839

\bibitem{creath92}
Creath K, Wyant J and Malacara E 1992 {\em Moiré and Fringe Projection
  Techniques\/} vol~2 (Wiley New York) pp 653--685

\bibitem{buckanie2013}
Buckanie N, Kirschbaum P, Sindermann S and {zu Heringdorf} F~J~M 2013 {\em
  Ultramicroscopy\/} {\bf 130} 49--53

\bibitem{tikuisis2017}
Tikui{\v s}is K~K, Beran L, Cejpek P, Uhl{\'i}{\v r}ov{\'a} K, Hamrle J, Va{\v
  n}atka M, Urb{\'a}nek M and Veis M 2017 {\em Materials \& Design\/} {\bf 114}
  31--39

\bibitem{ciesielski2017}
Ciesielski A, Skowronski L, Trzcinski M and Szoplik T 2017 {\em Applied Surface
  Science\/} {\bf 421} 349--356

\bibitem{temnov2010}
Temnov V~V, Armelles G, Woggon U, Guzatov D, Cebollada A, {Garcia-Martin} A,
  {Garcia-Martin} J~M, Thomay T, Leitenstorfer A and Bratschitsch R 2010 {\em
  Nature Photonics\/} {\bf 4} 107--111

\bibitem{zheng2015}
Zheng W, Liu X, Hanbicki A~T, Jonker B~T and L{\"u}pke G 2015 {\em Optical
  Materials Express\/} {\bf 5} 2597

\end{thebibliography}

\end{document}